\documentclass[aps,prl,twocolumn,showpacs,floatfix]{revtex4}
\usepackage{graphicx}
\usepackage{amsmath}
\begin{document}

\title{Current reversals in a rocking ratchet: dynamical vs symmetry-breaking 
mechanisms}

\author{D. Cubero$^{1}$, V. Lebedev\footnote{Present address:
Institut fu\"r Quantenoptik, Leibnitz Universit\"at Hannover, 
Welfengarten 1, D-30167 Hannover, Germany}$^{2}$ and F. Renzoni$^{2}$}

\affiliation{$^{1}$
Departamento de F\'{\i}sica Aplicada I, EUP, Universidad de Sevilla, 
Calle Virgen de \'Africa 7, 41011 Sevilla, Spain}

\affiliation{$^{2}$Department of Physics and Astronomy, University College
London, Gower Street, London WC1E 6BT, United Kingdom}

\date{\today}

\begin{abstract}
Directed transport in ratchets is determined by symmetry-breaking 
in a system out of equilibrium. A hallmark of rocking ratchets is 
current reversals: an increase in the rocking force changes the direction 
of the current. In this work for a bi-harmonically driven spatially symmetric
rocking ratchet we show that a class of current reversal is precisely 
determined by symmetry-breaking, thus creating a link between dynamical 
and symmetry-breaking mechanisms.
\end{abstract}


\maketitle

Many processes in physics, chemistry and biology involve directed transport 
through periodic structures.  For the equilibrium situation of Brownian
motion, diffusion can be turned into directed diffusion by the application
of a dc bias.  In out-of-equilibrium systems, new mechanisms for directed
transport may arise. Counter-intuitively, far from equilibrium it is possible 
to obtain directed transport through a macroscopically flat potential in 
the absence of an applied dc bias. This is the so-called ratchet effect 
\cite{comptes,magnasco,adjari,bartussek,doering,reimann,rmp09}.

The archetypal of a ratchet device is the {\it rocking ratchet}. In this setup,
Brownian particles experience an asymmetric sawtooth potential and a 
sinusoidal rocking force. The rocking force drives the system out of 
equilibrium, and directed transport is generated following the breaking of the 
symmetries of the system. An analogous effect can also be produced in 
a spatially symmetric potential and a biharmonic force, with the latter
playing the double role of driving the system out of equilibrium and breaking 
the relevant time-symmetries \cite{fabio,mahato,chialvo,dykman,goychuk,flach00,flach01,super,machura}.

A hallmark of rocking ratchets are current reversals. By progressively 
increasing the rocking force from zero, the generation of a current is observed,
whose magnitude is first an increasing function of the strength of the 
driving. However, at larger values of the rocking force the current reaches a 
maximum, then decreases to zero and changes sign. This feature can appear 
several times in a given system for different values of the force, thus
producing multiple current reversals. Single and multiple current reversals
have been observed in a variety of systems, both for an asymmetric potential
and a symmetric drive and for a symmetric potential and a time-asymmetric
drive \cite{jung,mateos,phil}.

Current reversals are usually considered a dynamical effect, not related to 
the symmetry-breaking required to allow directed motion. In this work, for 
the specific system with a spatially symmetric potential and a time-asymmetric
drive, we show that a class of current reversal is actually determined by 
dissipation-induced symmetry breaking. As a consequence, these reversals
are not present in the Hamiltonian limit, nor in the overdamped limit.

Our work consists into a theoretical analysis of the relationship between
current reversals and dissipation induced symmetry-breaking. This is carried
out comparing differing regimes: weakly damped, Hamiltonian, and overdamped.
In the case of weak damping, where current reversals associated to dissipative
effects are present, the theoretical analysis is also supported by experimental 
results obtained with cold atom ratchets. 

In our theoretical analysis, the dynamics of particles in the considered 
spatially symmetric rocking ratchet is described by the Langevin equation
\begin{equation}
m\ddot{x} = -\alpha\dot{x}-U'(x)+F(t)+\xi (t),
\end{equation}
where $U(x)=U_0\cos(2kx)/2$ is a periodic potential,  $\alpha$ the friction
coefficient, $\xi(r)$ a Gaussian white noise:
$\langle \xi(t)\rangle =0$, $\langle \xi(t)\xi(t')\rangle =2D\delta(t-t')$,
and $F(t)$ is an applied bi-harmonic drive of the form \footnote{To avoid a 
dependence of the current with the initial conditions, the bi-harmonic driving 
is switched on adiabatically both in the simulations and the experiments.}
\begin{equation}
F(t) = F_0\left[ A \cos(\omega t) + B \cos(2\omega t +\phi)\right].
\label{eq:drive}
\end{equation}

The generation of a current in such out-of-equilibrium set-up can be 
understood within the framework of the symmetry analysis 
\cite{flach00,flach01,super}.
For the considered spatially symmetric potential there are two 
time-symmetries which need to be broken to allow for the generation of
a current: the shift symmetry, which corresponds to invariance under 
the transformation $(x,p,t)\to (-x,-p,t+T/2)$, with $T$ the period of the 
drive, and the time-reversal symmetry, which requires invariance under 
the transformation $(x,p,t)\to (x,-p,-t)$. For a bi-harmonic drive of
the form of Eq.~\ref{eq:drive}, the shift symmetry is broken independently 
of the value of $\phi$ (for$A\neq 0$, $B\neq 0$). The breaking of the 
time-reversal symmetry depends both on the value of the phase $\phi$
and the dissipation level. In the Hamiltonian (dissipationless) case,
the system is invariant under time reversal for $\phi=n\pi$, with $n$ 
integer. Thus, for these values of the relative phase, no current can be
generated. However, for nonzero, weak dissipation the time-reversal
symmetry is broken by dissipation, and directed motion can be produced 
also for $\phi=n\pi$. In the regime of weak dissipation, the dependence of
the particles' velocity $v$ on the phase $\phi$ is well described, in 
leading order, by $v=v_{\rm max}\sin(\phi-\phi_0)$ where $\phi_0$ is determined
by dissipation, and vanishes in the Hamiltonian limit. It has been shown 
recently that this sine-like functional form is a consequence solely of 
the system symmetries, being independent of the interaction details \cite{niurka}.  
Of importance for the present study is the overdamped regime. 
In this limit, the so-called "supersymmetry" \cite{super} $(x,p,t)\to (x+\lambda/2,-p,-t)$,
with $\lambda$ the spatial period of the potential, is satisfied for 
$\phi=\pi/2 + n\pi$, with $n$ integer. For these values of the driving 
phase $\phi$ no directed transport can occur.

In order to establish a link between current reversals and symmetry breaking,
we examine the dependence of the particles' current on two different
quantities.  First, we study the current as a function of the driving phase
$\phi$, which controls the time-symmetry of the Hamiltonian. This allows us
to reveal the role of dissipation-induced symmetry breaking. Second, we 
consider the standard set-up for the observation of current-reversals:
we fix the Hamiltonian by choosing a value of $\phi$ which corresponds 
to broken time-reversal symmetry ($\phi=\pi/2$, say) and study the current
as a function of the driving strength. This will allow us to detect current
reversals, and relate them to dissipation-induced symmetry breaking. In all the
results presented in this work, the relative weight between the harmonic of the
force is fixed ($A=B=1$ in all numerical simulations) and we vary the overall
amplitude $F_0$.

\begin{figure}[h]
\begin{center}
\includegraphics[height=2.2in]{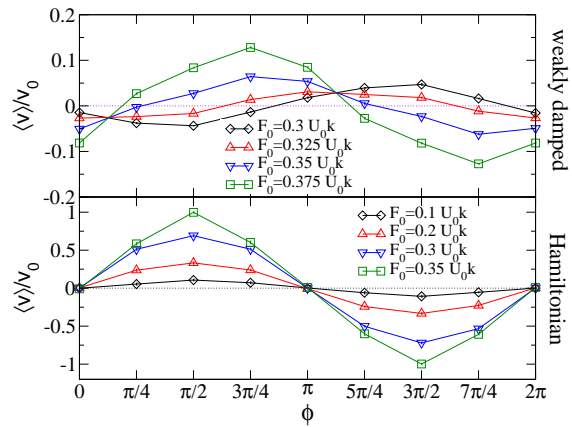}
\end{center}
\caption{(Color online) Average atomic velocity as a function of the relative
phase $\phi$ between harmonics of the ac drive for several values of the driving
amplitude $F_0$ and $\omega=\omega_v=k(2U_0/m)^{1/2}$.
Top panel: simulation data in the weakly damped regime. The friction and the noise
strength values are fixed to $\alpha=0.15 \alpha_0$ and $D=1.944 D_0$, respectively, 
where $\alpha_0=m k v_0$, $D_0=\alpha_0^2 v_0/k$, and $v_0=(U_0/m)^{1/2}/10$. 
The values near $\phi=\pi/2$ (or $\phi=3\pi/2$) show a current reversal as the 
driving amplitude is increased.  Bottom panel: simulation data in the Hamiltonian regime 
($\alpha=D=0$). The initial conditions were chosen within the chaotic sea shown in 
Fig.~\ref{fig:fig3}. Lines are a guide to the eye.}
\label{fig:fig1}
\end{figure}

We examine first the weakly damped regime. The top panel of Figure \ref{fig:fig1} shows the 
average particles' velocity as a function of the driving phase for different
values of the amplitude of the drive. The displacement of the current curves' 
extrema clearly indicates that, for a given dissipation level (i.e., a given value 
of the friction coefficient $\alpha$), a variation 
of the drive amplitude leads to a variation in the dissipation-induced phase
lag $\phi_0$.  This is the central point of our analysis: because of 
dissipation, the curve of the average velocity vs the phase $\phi$ acquires
a nonzero phase lag $\phi_0$, and the magnitude of the phase lag is a function 
of the strength of the driving.  If we now examine the behavior of the 
current for a fixed phase $\phi$ (say, $\phi=\pi/2$) we see that the variation
of $\phi_0$ due to the change in driving strength leads to a current reversal.
Thus, a link between current reversals and dissipation-induced symmetry 
breaking is established.

Our argument relies on the existence of a dissipation-induced symmetry breaking
phase lag $\phi_0$, whose value depends on the dissipation level and driving 
strength. The consistency of the argument can be verified by considering two
extreme limits: the Hamiltonian case, and the overdamped regime. In both cases,
the phase lag $\phi_0$ is "locked" to a given value ($0$ for the Hamiltonian
case and $\pi/2$ in the overdamped regime) by the system symmetry, and cannot
be varied by modifying the driving strength. According to our argument, current 
reversals should disappear in  both limits.

We consider the dissipationless limit. The bottom panel of Figure \ref{fig:fig1} shows the 
average particles' velocity as a function of the phase $\phi$ for different
values of the driving strength. As in the weakly damped case, the current
shows a sine-like dependence on the phase $\phi$. However, unlike the 
weakly-damped case, the zeros of the current curves are now fixed by the 
time-reversal symmetry, and a change in the driving strength does not induce
any phase shift. If we now consider the current dependence on the driving 
strength $F_0$ for a given phase (say, $\phi=\pi/2$), we observe that no 
current reversal occurs. These results are consistent with the link we 
established between current reversals and dissipation-induced symmetry 
breaking.


\begin{figure}[h]
\begin{center}
\includegraphics[height=2.in]{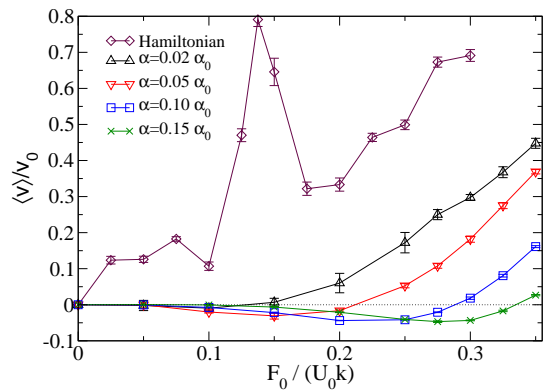}
\end{center}
\caption{(Color online) 
Simulation data in the weakly damped regime: Current as a function of the 
driving amplitude $F_0$ for several values of the friction $\alpha$ and a 
fixed driving phase $\phi=\pi/2$. Rest of the parameters as in Fig.~\ref{fig:fig1}. 
As the dissipation is decreased, the current reversal position is shifted to lower 
values of the driving amplitude $F_0$. The diamonds show the results for the Hamiltonian 
system. Lines are a guide to the eye.
}
\label{fig:fig2}
\end{figure}

\begin{figure}[hbtp]
\begin{center}
\includegraphics[height=2.in]{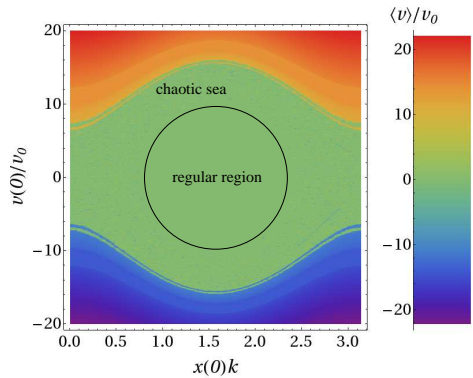}
\end{center}
\caption{(Color online) 
Simulation data in the Hamiltonian regime: The current as a function of the 
initial conditions $v(0)$ and $x(0)$ is represented using a color density plot. 
The driving parameters are $\omega=\omega_v$, $F_0=0.1 U_0k$ and $\phi=\pi/2$. 
A black circumference indicates the boundary of a regular circular region in
which the current is zero throughout.
}
\label{fig:fig3}
\end{figure}

\begin{figure}[hbtp]

\begin{center}
\includegraphics[height=2.8in]{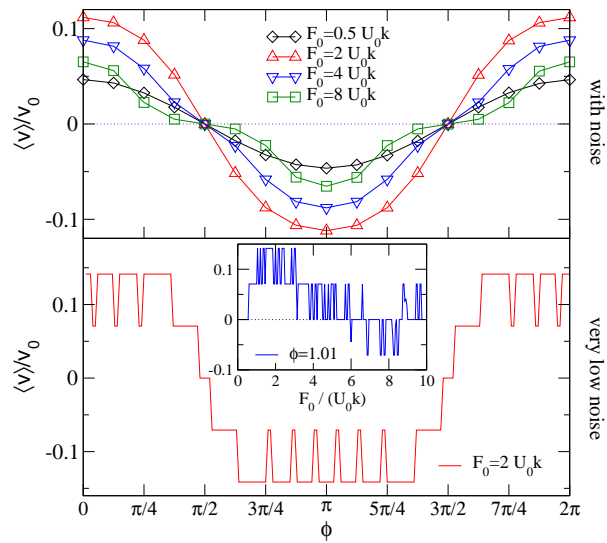}
\end{center}
\caption{(Color online) 
Simulation data in the overdamped regime: Current as a function of the 
driving phase $\phi$ for several values of the driving amplitude $F_0$, 
with $\omega=0.01\omega_v$ and a friction $\alpha=100\alpha_0$. Top panel: 
$D=0.1944 D_0$, being the lines a guide to the eye. Bottom panel: 
$D=0.1944\cdot10^{-3} D_0$. The inset in the bottom panel shows the current 
as a function of $F_0$ for a fixed  driving phase $\phi=1.01$, displaying 
multiple current reversals.
}
\label{fig:fig4}
\end{figure}

It is interesting to study the dynamics of disappearance of current reversals
while approaching the Hamiltonian limit. This is done in Fig.~\ref{fig:fig2},
where  for a fixed phase $\phi=\pi/2$ the current is studied as a function
of the driving strength, for different levels of dissipation. It appears 
that by decreasing dissipation, the position of the current reversal 
moves towards $F_0=0$, and in the Hamiltonian limit the current reversal 
disappears as its position coincides with $F_0=0$.

We notice that in the Hamiltonian limit the asymptotic 
particles' velocity depends on the initial preparation \cite{sum_rule}. For
the specific system of interest here, such a dependence is summarized in
Fig.~\ref{fig:fig3}. There, we also evidence the chaotic sea. It is within that
region that the initial condition for the calculations in the Hamiltonian
limit presented here (bottom panel of Fig. \ref{fig:fig1} and one set in Fig.\ref{fig:fig2})
were chosen.

We now consider the overdamped case. Numerical simulations for this regime, 
in the presence of noise, are reported in the top panel of Fig. \ref{fig:fig4}. 
The current is still described by a sine-like function $v\sim\sin (\phi-\phi_0)$
but the value of the phase lag is now locked by the symmetry to $\phi_0=\pi/2$. A 
variation in the strength of the driving does not lead to a change in the
phase lag.  Studying the current as a function of the driving strength, for
a fixed value $\phi=\pi/2$ of the driving phase, reveals the absence of 
current reversals. This confirms our statement about a link between current
reversals and the symmetry breaking induced by the presence of a moderate 
amount of dissipation.

\begin{figure}[h]
\begin{center}
\includegraphics[height=4.5in]{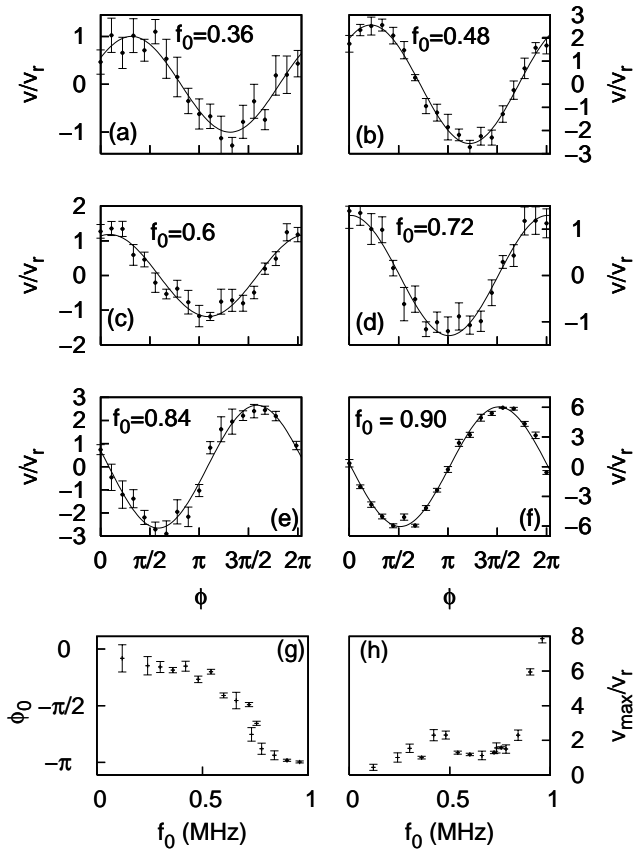}
\end{center}
\caption{Experimental results for 1D rocking ratchet for cold
atoms. (a)-(f): average atomic velocity, rescaled by the recoil velocity
$v_r$ ($v_r = 5.88 $ mm/s for $^{87}$Rb), as a function of the relative phase 
$\phi$ between harmonics of the ac drive, for different values of the driving 
force amplitude. (g) Dissipation-induced phase lag $\phi_0$, as obtained by 
fitting data as those in (a)-(f) with the function $v/v_r=A\sin(\phi-\phi_0)$,
as a function of the driving amplitude. The amplitude $A$ is reported 
in (h). The parameters of the optical lattice are: detuning from resonance
$\Delta = -15 \Gamma$ and intensity per lattice beam $I_L= 105$ mW/cm$^2$. The
rocking force is of the form of Eq.~\protect\ref{eq:drive}, with 
$\omega/(2\pi)=100$ kHz, $A=1$, $B=2$ and $F_0 = - mf_0\omega/k$ where $m$
is the atomic mass,  $k$ the laser field wave-vector, and the value for 
$f_0$ (in MHz) for the different set of data is reported in the figures.}
\label{fig:fig5}
\end{figure}

So far we have demonstrated that the dependence of the dissipation-induced
phase lag $\phi_0$  on the driving strength $F_0$ results in current
reversals, as observed by monitoring the dependence of the current at a
fixed driving phase on the strength of the drive. For the same argument, this
type of current reversals are absent in the Hamiltonian and overdamped regime,
where the phase lag $\phi_0$ does not vary  with the amplitude of the drive.
It is important to specify the regime of applicability of such a reasoning,
and underline that not all current reversals may be necessarily traced back
to dissipation-induced symmetry breaking. In fact our argument relies on the
assumption that the current can be well described by a smooth sine-like
function $v\sim\sin (\phi-\phi_0)$. This holds provided that the noise level 
is sufficiently large to smooth the curve. Thus, at low noise levels additional
current reversals may appear, not related to dissipation-induced symmetry
breaking. An example of this is provided in the bottom panel of Fig.~\ref{fig:fig4}, 
where current reversals are observed in the overdamped regime at a very 
low level of noise.

The weakly damped regime, of central interest here, can be explored using
dissipative cold atoms ratchets \cite{brown}. We used a 1D rocking ratchet
set-up for $^{87}$Rb atoms \cite{advances}, which corresponds to a spatially 
symmetric potential and a bi-harmonic rocking force of the form of Eq.~\ref{eq:drive}.
Proceeding along the lines of the theoretical analysis, we measured the atomic
average velocity as a function of the driving phase $\phi$, for a given
dissipation level and for different strengths of the driving. Our results,
shown in Fig.~\ref{fig:fig5}, confirm the prediction of the general theory we
presented. An increase in the applied driving force amplitude $F_0$ leads
to a large variation in the phase lag $\phi_0$, as also summarized in
Fig.~\ref{fig:fig5}(g). Also the amplitude of the curve varies with the
driving amplitude $F_0$. However, it never becomes zero (see
Fig.~\ref{fig:fig5}(h)). It is thus the  variation in the phase lag $\phi_0$
which produces a change in sign of the current observed for $\phi=\pi/2$, i.e.
a current reversal.

In conclusion, in this work we studied, both theoretically and experimentally,
the relationship between current reversals and symmetry-breaking. For
the specific system with a spatially symmetric potential and a time-asymmetric
drive, we showed that a class of current reversal is actually determined by
dissipation-induced symmetry breaking. As a consequence, these reversals
are not present in the Hamiltonian limit, nor in the overdamped limit.

This research was supported by the Leverhulme Trust. One of us (DC)
also thanks the Ministerio de Ciencia e Innovaci\'on of Spain for financial
support (grant FIS2008-02873).

\end{document}